\documentclass[12pt,english]{article}
\usepackage{ae,aecompl}

\usepackage[T1]{fontenc}
\usepackage[latin9]{inputenc}
\usepackage{geometry}
\usepackage{babel}
\usepackage{amsmath}
\usepackage{amssymb}
\usepackage{esint}
\usepackage[unicode=true]
 {hyperref}
\usepackage{breakurl}

\makeatletter
\numberwithin{equation}{section}

\usepackage{babel}

\makeatother

\begin{document}
\begin{flushright}
FIAN/TD/2017-15\\
 
\par\end{flushright}

\vspace{0.5cm}

\begin{center}
\textbf{\large{}On current contribution to Fronsdal equations}
\par\end{center}{\large \par}

\begin{center}
\vspace{0.5cm}

\par\end{center}

\begin{center}
\textbf{N.G.~Misuna}\\
 \vspace{0.5cm}
 \textit{I.E. Tamm Department of Theoretical Physics, Lebedev Physical
Institute,}\\
 \textit{ Leninsky prospect 53, 119991, Moscow, Russia}\\

\par\end{center}

\begin{center}
\textit{Moscow Institute of Physics and Technology,}\\
 \textit{ Institutsky lane 9, 141700, Dolgoprudny, Moscow region,
Russia}\\

\par\end{center}

\begin{center}
\vspace{0.5cm}

\par\end{center}
\begin{abstract}
\noindent We explore a local form of second-order Vasiliev equations
proposed in {[}arXiv:1706.03718{]} and obtain an explicit expression
for quadratic corrections to bosonic Fronsdal equations, generated
by gauge-invariant higher-spin currents. Our analysis is performed
for general phase factor, and for the case of parity-invariant theory
we find the agreement with expressions for cubic vertices available
in the literature. This provides an additional indication that local
frame proposed in {[}arXiv:1706.03718{]} is the proper one.\newpage{}
\end{abstract}

\section{Introduction}

Linear equations that describe a free propagation of massless higher-spin
(HS) fields were found a long time ago by Fronsdal and Fang \cite{Fronsdal,Fang Fronsdal}.
But to build any consistent nonlinear deformation of them turned out
to be extremely nontrivial task. Up to date the only available example
of full nonlinear HS gauge theory is provided by Vasiliev equations
\cite{Vas_eq90,Vas_eq92}. They represent an interacting theory of
massless fields of all spins over anti-de Sitter ($AdS$) background.
As opposed to e.o.m. of standard field theory, Vasiliev equations
are so-called unfolded ones, i.e. they represent first-order differential
equations in terms of exterior (0- and 1-) forms. Each field of given
spin is described by an infinite number of unfolded fields parametrising
all its degrees of freedom. An infinite number of vertices, describing
HS interactions, are encoded into the evolution over auxiliary twistor-like
variables. In this regard Vasiliev equations can be considered as
generating ones (``equations for equations'').

A reconstruction of space-time dynamics from Vasiliev equations is
a nontrivial problem, essentially because of the freedom in the choice
of resolution operator for twistor-like variables. As usual, the resolution
operator is determined up to an arbitrary solution of homogeneous
equation, that in terms of physical fields amounts to the freedom
in a field redefinition, which can affect the form of e.o.m. For example,
in \cite{ProkVas} it was found that by nonlocal field redefinitions
one can get rid of interactions in $3d$ HS equations (see also \cite{ProkVas_coh,KesGomSkvTar}
for the proof of pseudolocal-triviality of any $3d$ HS currents).
In \cite{BoulKesSkvTar} it was shown that the simplest choice of
the resolution operator lead to nonlocal expressions for $4d$ cubic
HS vertices. All that brings up a question of admissible functional
class of field redefinitions \cite{Vas_Loc,SkvTar,Tar_Proc,Tar_Redef}.
On the other hand, field redefinitions, bringing quadratic equations
to the local form, were found in \cite{Vas_0form} for the sector
of 0-forms and in \cite{GelVas_1form} for the sector of 1-forms.
These were tested in \cite{DidVas_3pnt,Tar_unpubl,SezgSkvZhu_3pnt},
where it was shown that the resulting local HS equations properly
reproduce holographic correlators in accordance with Klebanov--Polyakov
HS $AdS/CFT$ conjecture \cite{KlebPol}. Later, in \cite{Vas_resol}
it was shown how to construct a proper resolution operator, enforcing
the locality at the second order and minimising nonlocality at higher
orders. Formally this operator can be considered as the resolution
operator of \cite{BoulKesSkvTar}, rectified by non-local field redefinitions
of \cite{Vas_0form,GelVas_1form}.

In this note we provide a further analysis of unfolded local quadratic
equations of \cite{GelVas_1form} and obtain an explicit form of corrections
to bosonic Fronsdal equations that are generated by gauge-invariant
HS currents. These should be compared with results of \cite{Metsaev}
where HS cubic couplings were found in flat space in lightcone formulation,
and \cite{SltTar} where they were restored via $AdS/CFT$ from correlators
of boundary free scalar theory and later in \cite{SlTar_Noether}
shown to solve the bulk Noether procedure. Expressions for quadratic
corrections we found turn out to be in the full agreement with these
results, thus providing one more confirmation that the local frame
of \cite{Vas_0form,GelVas_1form} is the appropriate one. In addition,
we worked out the dependence of vertices on the phase factor entering
Vasiliev equations, thus extending previous results to parity-breaking
theories. It turns out that there is a specific value of the phase
$\varphi=\tfrac{\pi}{4}$, where leading-derivative vertex maximally
breaks parity, which may have interesting implications for dual boundary
theory.

\section{Higher-Spin Equations}

HS equations in four dimensions are \cite{Vas_eq92}
\begin{eqnarray}
 &  & \mathrm{d}W+W*\land W=-i\theta_{\alpha}\land\theta^{\alpha}\left(1+\eta B*\varkappa k\right)-i\bar{\theta}_{\dot{\alpha}}\land\bar{\theta}^{\dot{\alpha}}\left(1+\bar{\eta}B*\bar{\varkappa}\bar{k}\right),\label{HS_1}\\
 &  & \mathrm{d}B+W*B-B*W=0.\label{HS_2}
\end{eqnarray}
Here $\mathrm{d}$ is the space-time de Rham differential, $W$ and
$B$ are master-fields of the theory (onwards we omit wedge symbol)
dependent on space-time coordinates and twistor-like variables $Y^{A}=\left(y^{\alpha},\bar{y}^{\dot{\alpha}}\right)$,
$Z^{A}=\left(z^{\alpha},\bar{z}^{\dot{\alpha}}\right)$ with two-valued
spinor indices $\alpha$ and $\dot{\alpha}$. The $Y$ and $Z$ realise
the HS algebra through the noncommutative star product
\begin{equation}
(f*g)(Z,Y)=\int d^{4}Ud^{4}Ve^{iU_{A}V^{A}}f(Z+U,Y+U)g(Z-V,Y+V),
\end{equation}
with the integration measure fixed so as $1*f=f*1=1$. Spinor indices
are raised and lowered via $sp\left(2\right)$-metrics
\begin{equation}
v^{\alpha}=\epsilon^{\alpha\beta}v_{\beta},\quad v_{\alpha}=\epsilon_{\beta\alpha}v^{\beta},\quad\bar{v}^{\dot{\alpha}}=\epsilon^{\dot{\alpha}\dot{\beta}}\bar{v}_{\dot{\beta}},\quad\bar{v}_{\dot{\alpha}}=\epsilon_{\dot{\beta}\dot{\alpha}}\bar{v}^{\dot{\beta}}.
\end{equation}
$sp\left(4\right)$-indices are transformed by $\epsilon_{AB}$ built
from $\epsilon_{\alpha\beta}$ and $\epsilon_{\dot{\alpha}\dot{\beta}}$

\begin{equation}
V^{A}=\epsilon^{AB}V_{B},\quad V_{A}=\epsilon_{BA}V^{B}.
\end{equation}
$\varkappa$ and $\bar{\varkappa}$ in (\ref{HS_1}) are inner Klein
operators, which are specific elements of the star-product algebra
\begin{equation}
\varkappa:=\exp\left(iz_{\alpha}y^{\alpha}\right),\quad\bar{\varkappa}:=\exp\left(i\bar{z}_{\dot{\alpha}}\bar{y}^{\dot{\alpha}}\right),
\end{equation}
having the distinguishing properties

\begin{equation}
\varkappa*\varkappa=1,\quad\varkappa*f\left(z^{\alpha},y^{\alpha}\right)=f\left(-z^{\alpha},-y^{\alpha}\right)*\varkappa,
\end{equation}

\begin{equation}
f\left(z,y\right)*\varkappa=f\left(-y,-z\right)e^{iz_{\alpha}y^{\alpha}},
\end{equation}
and analogously for $\bar{\varkappa}$.

Master-field $B$ is a 0-form, while $W$ is a 1-form in a space-time
differential $dx^{\underline{m}}$ or in an auxiliary differential
$\theta^{A}$ dual to $Z^{A}$. All differentials anticommute

\begin{equation}
\left\{ dx^{\underline{m}},dx^{\underline{n}}\right\} =\left\{ dx^{\underline{m}},\theta^{A}\right\} =\left\{ \theta^{A},\theta^{B}\right\} =0.
\end{equation}
Besides the inner Klein operators there is also a pair of exterior
Klein operators $K=\left(k,\bar{k}\right)$ which have similar properties
to $\left(\varkappa,\bar{\varkappa}\right)$ 
\begin{equation}
kk=1,\quad kf(z^{\alpha};y^{\alpha};\theta^{\alpha})=f(-z^{\alpha};-y^{\alpha};-\theta^{\alpha})k,\label{extKlein}
\end{equation}
(analogously for $\bar{k}$), but $k$ ($\bar{k}$) in addition anticommute
with $\theta$ ($\bar{\theta}$) differentials that does not permit
to realise them as elements of the star-product algebra.

Thus the full arguments of master-fields are
\begin{equation}
W=W\left(Z;Y|K|x|\theta^{A},dx^{\underline{m}}\right),\quad B=B\left(Z;Y|K|x\right).
\end{equation}
$K$-dependence of the fields leads to the splitting of the field
spectrum into topological and physical sectors. The first one describes
finite-dimensional modules and contains $W$ linear in $k$ or $\bar{k}$
and $B$ depending on $k\bar{k}$. We truncate it away. The physical
sector describing relativistic fields contains $W$ depending on $k\bar{k}$
and $B$ linear in $k$ or $\bar{k}$. Moreover, in this note we consider
a bosonic reduction, which leaves only one field of every integer
spin and is reached by setting
\begin{equation}
W\left(Z;Y|K|x|\theta^{A},dx^{\underline{m}}\right)\rightarrow W\left(Z;Y|x|\theta^{A},dx^{\underline{m}}\right)\left(1+k\bar{k}\right),\quad B\left(Z;Y|K|x\right)\rightarrow B\left(Z;Y|x\right)\left(k+\bar{k}\right).
\end{equation}
$\eta$ in (\ref{HS_1}) is a free complex parameter of the theory
which can be normalised to be unimodular\footnote{In \cite{Vas_eq92} it was conjectured that a different situation
when $\left(\eta\right)$ $\bar{\eta}=0$ corresponds to (anti)selfdual
HS theory, allowing no nontrivial amplitudes. Here we do not consider
this case.} $\eta\bar{\eta}=1$, hence representing the phase factor freedom.
HS theory is parity-invariant in the two cases of $\eta=1$ (A-model)
and $\eta=i$ (B-model) \cite{SezgSun}.

\section{Perturbation theory}

To start a perturbative expansion one has to fix some vacuum solution
to (\ref{HS_1}), (\ref{HS_2}). Eq. (\ref{HS_2}) can be solved by
setting the vacuum value of $B$ to zero
\begin{equation}
B_{0}=0.\label{B_0}
\end{equation}
Then the solution for (\ref{HS_1}) can be chosen as
\begin{equation}
W_{0}=\omega_{AdS}+Z_{A}\theta^{A},\label{W_0}
\end{equation}
with the space-time 1-form of $sp(4)$-connection $\omega_{AdS}$
describing the $AdS_{4}$ background 
\begin{eqnarray}
 &  & \omega_{AdS}=-\dfrac{i}{4}\left(\omega_{L}^{\alpha\beta}y_{\alpha}y_{\beta}+\bar{\omega}_{L}^{\dot{\alpha}\dot{\beta}}\bar{y}_{\dot{\alpha}}\bar{y}_{\dot{\beta}}+2\lambda h^{\alpha\dot{\beta}}y_{\alpha}\bar{y}_{\dot{\beta}}\right),\\
 &  & \mathrm{d}\omega_{AdS}+\omega_{AdS}*\omega_{AdS}=0,
\end{eqnarray}
where $\lambda$ is the cosmological parameter (inverse radius of
$AdS$).

Performing an expansion of (\ref{HS_1})-(\ref{HS_2}) around vacuum
(\ref{B_0})-(\ref{W_0}) one gets at the linear order
\begin{eqnarray}
 &  & \mathcal{D}_{ad}\omega\left(Y|K|x\right)=L\left(C\right),\label{On_sh_Th_1}\\
 &  & \mathcal{D}_{tw}C\left(Y|K|x\right)=0,\label{On_sh_Th_2}
\end{eqnarray}
where
\begin{eqnarray}
 &  & L\left(C\right):=\dfrac{i\lambda}{4}\eta\bar{H}^{\dot{\alpha}\dot{\beta}}\bar{\partial}_{\dot{\alpha}}\bar{\partial}_{\dot{\beta}}C\left(0,\bar{y}|K|x\right)k+\dfrac{i\lambda}{4}\bar{\eta}H^{\alpha\beta}\partial_{\alpha}\partial_{\beta}C\left(y,0|K|x\right)\bar{k},\\
 &  & H^{\alpha\beta}:=h^{\alpha\dot{\gamma}}h^{\beta}\phantom{}_{\dot{\gamma}},\quad\bar{H}^{\dot{\alpha}\dot{\beta}}:=h^{\gamma\dot{\alpha}}h_{\gamma}\phantom{}^{\dot{\beta}}\\
 &  & \partial_{\alpha}:=\dfrac{\partial}{\partial y^{\alpha}},\quad\bar{\partial}_{\dot{\alpha}}:=\dfrac{\partial}{\partial\bar{y}^{\dot{\alpha}}},\\
 &  & \mathcal{D}_{ad}f\left(Y|K|x\right):=D^{L}f+\lambda h^{\alpha\dot{\beta}}\left(y_{\alpha}\bar{\partial}_{\dot{\beta}}+\partial_{\alpha}\bar{y}_{\dot{\beta}}\right)f,\label{D_ad}\\
 &  & \mathcal{D}_{tw}f\left(Y|K|x\right):=D^{L}f-i\lambda h^{\alpha\dot{\beta}}\left(y_{\alpha}\bar{y}_{\dot{\beta}}-\partial_{\alpha}\bar{\partial}_{\dot{\beta}}\right)f,\label{D_tw}\\
 &  & D^{L}f:=\mathrm{d}f+\left(\omega_{L}^{\alpha\beta}y_{\alpha}\partial_{\beta}+\bar{\omega}_{L}^{\dot{\alpha}\dot{\beta}}\bar{y}_{\dot{\alpha}}\bar{\partial}_{\dot{\beta}}\right)f.
\end{eqnarray}
Eqs. (\ref{On_sh_Th_1})-(\ref{On_sh_Th_2}) represent a so-called
unfolded form of Fronsdal equations, describing free propagation of
HS fields over $AdS_{4}$ background. Let us expand HS fields as
\begin{equation}
\omega\left(Y|K|x\right)=\sum_{m,n=0}^{\infty}\omega_{m,n}\left(Y|K|x\right),\quad C\left(Y|K|x\right)=\sum_{m,n=0}^{\infty}C_{m,n}\left(Y|K|x\right),
\end{equation}
where
\begin{equation}
f_{m,n}\left(Y\right):=f_{\alpha_{1}...\alpha_{m},\dot{\beta}_{1}...\dot{\beta}_{n}}y^{\alpha_{1}}...y^{\alpha_{m}}\bar{y}^{\dot{\beta}_{1}}...\bar{y}^{\dot{\beta}_{n}}.
\end{equation}
We also introduce a decomposition into different helicity-sign sectors
as
\begin{equation}
\omega=\omega_{+}+\omega_{-}+\omega_{0},\quad C=C_{+}+C_{-}+C_{0},
\end{equation}
where
\begin{equation}
f_{+}=\sum_{m>n}f_{m,n},\quad f_{-}=\sum_{m<n}f_{m,n},\quad f_{0}=\sum_{m=n}f_{m,n}
\end{equation}
are positive-helicity, negative-helicity and zero-helicity sectors
respectively. Then a submodule describing spin-$s$ field consists
of $\omega_{m,n}$, $n+m=2\left(s-1\right)$ and $C_{m,n}$, $\left|m-n\right|=2s$.

At the second order one should make field redefinitions that brings
equations to the local frame, removing infinite higher-derivative
tails. Such redefinitions were found in \cite{Vas_0form,GelVas_1form}.
Applying them one obtains
\begin{eqnarray}
 &  & \mathcal{D}_{ad}\omega+\left[\omega,\omega\right]_{*}=L\left(C\right)+Q\left(C,\omega\right)+\Gamma_{s<s_{1}+s_{2}}\left(J\right)+\Gamma^{can}\left(J\right),\label{HS_cur_1}\\
 &  & \mathcal{D}_{tw}C\left(Y|K|x\right)+\left[\omega,C\right]_{*}=-\mathcal{H}_{\eta}\left(J\right)-\mathcal{H}_{\bar{\eta}}\left(J\right)+\mathcal{D}_{tw}B^{sum}\left(J\right),\label{HS_cur_2}
\end{eqnarray}
where 
\begin{equation}
J\left(Y^{1},Y^{2}|K|x\right):=C\left(Y^{1}|K|x\right)C\left(Y^{2}|K|x\right)
\end{equation}
is a bilinear HS current. The above-mentioned redefinitions serve
to make $J$-dependent terms local. We will analyse the first equation
(\ref{HS_cur_1}) that comprise Fronsdal equations with quadratic
corrections. These corrections are of the four types: $\left[\omega,\omega\right]_{*}$
term which is completely fixed by HS symmetry algebra; gauge-dependent
contribution $Q\left(C,\omega\right)$ which is local from the very
beginning because $\omega$ is a polynomial in $Y$ of restricted
degree for any fixed spin; $\Gamma_{s<s_{1}+s_{2}}\left(J\right)$
being the current deformation in gauge-dependent sector inside the
triangle inequality $s<s_{1}+s_{2}$; $\Gamma^{can}\left(J\right)$
which is gauge-invariant current deformation outside the triangle
inequality, $s\geq s_{1}+s_{2}$. It is this last contribution that
we are interested in.

Now we convert all objects to 0-forms expanding them in terms of vierbeins
\begin{equation}
\omega_{m,n}=h^{\alpha\dot{\beta}}\omega_{m,n|\alpha\dot{\beta}},\quad D^{L}=h^{\alpha\dot{\beta}}D_{\alpha\dot{\beta}}.
\end{equation}
Then one can rewrite a relevant sector of (\ref{HS_cur_1}) describing
current contribution to spin-$s$ field e.o.m. as \cite{GelVas_1form}
\begin{eqnarray}
 &  & D_{\alpha\dot{\beta}}\omega_{s-2,s|\alpha}\phantom{}^{\dot{\beta}}=-\bar{y}_{\dot{\beta}}\partial_{\alpha}\omega_{s-1,s-1|\alpha}\phantom{}^{\dot{\beta}}-y_{\alpha}\bar{\partial}_{\dot{\beta}}\omega_{s-3,s+1|\alpha}\phantom{}^{\dot{\beta}}+\partial_{\alpha}\partial_{\alpha}\mathcal{J}_{s,s},\label{unf_1}\\
 &  & D_{\beta\dot{\alpha}}\omega_{s,s-2|}\phantom{}^{\beta}\phantom{}_{\dot{\alpha}}=-y_{\beta}\bar{\partial}_{\dot{\alpha}}\omega_{s-1,s-1|}\phantom{}^{\beta}\phantom{}_{\dot{\alpha}}-\bar{y}_{\dot{\alpha}}\partial_{\beta}\omega_{s+1,s-3|}\phantom{}^{\beta}\phantom{}_{\dot{\alpha}}+\bar{\partial}_{\dot{\alpha}}\bar{\partial}_{\dot{\alpha}}\mathcal{J}_{s,s},\label{unf_2}
\end{eqnarray}
where
\begin{eqnarray}
\mathcal{J}_{s,s} & = & i\dfrac{\left(s-2\right)!}{8\left(2s\right)!}\sum_{k,m=0}^{s}\dfrac{\left(m+k\right)!\left(2s-m-k\right)!}{\left(s-k\right)!k!\left(s-m\right)!m!}\left(y^{\alpha}\partial_{\alpha}^{1}\right)^{m}\left(-y^{\beta}\partial_{\beta}^{2}\right)^{s-m}\left(\bar{y}^{\dot{\alpha}}\bar{\partial}_{\dot{\alpha}}^{1}\right)^{s-k}\left(-\bar{y}^{\dot{\beta}}\bar{\partial}_{\dot{\beta}}^{2}\right)^{k}\nonumber \\
 &  & \left\{ \sum_{n=0}^{s}\dfrac{i^{n}}{\left(s+n-1\right)!}\left(\left(\partial_{\gamma}^{1}\partial^{2\gamma}\right)^{n}+\left(\bar{\partial}_{\dot{\gamma}}^{1}\bar{\partial}^{2\dot{\gamma}}\right)^{n}\right)C\left(Y^{1}|K|x\right)C\left(Y^{2}|K|x\right)\right\} \Bigr|_{Y^{1}=Y^{2}=0}.\label{J_s,s}
\end{eqnarray}

\section{Currents contribution to Fronsdal equations}

Our goal is to develop an explicit expression for quadratic corrections
to Fronsdal equations that are generated by (\ref{unf_1})-(\ref{unf_2}).
Double-traceless field of spin-$s$ is described in terms of spinors
as $\omega_{\alpha\left(s-1\right),\dot{\alpha}\left(s-1\right)|\beta\dot{\beta}}$.
We make use of the fact that the currents in question are conformal
\cite{GelVas_cnfrml}, so we can keep track only of totally traceless
(in Lorentz tensor language) components of the Fronsdal fields, that
in spinor language corresponds to totally symmetric spinor-tensors
$\phi_{\alpha\left(s\right),\dot{\alpha}\left(s\right)}$ 
\begin{equation}
\phi_{s,s}:=\omega_{s-1,s-1|\beta\dot{\beta}}y^{\beta}\bar{y}^{\dot{\beta}}.
\end{equation}
Next, as we are on-shell we can take our fields to be transverse
\begin{equation}
D_{\alpha\dot{\beta}}\partial^{\alpha}\bar{\partial}^{\dot{\beta}}\phi_{s,s}=0.
\end{equation}
Finally, an important fact is that although the full nonlinear HS
theory does not admit a flat limit, cubic couplings we are studying
do admit it (for Fradkin-Vasiliev $2-s-s$ vertex \cite{FradVas1,FradVas2}
this was shown in \cite{Boul_Lecl_Sund}; see also \cite{JngTar}).
So we can take a flat limit in our equations and consider derivatives
to be commuting 
\begin{equation}
\left[D_{\alpha\dot{\alpha}},D_{\beta\dot{\beta}}\right]=0.
\end{equation}
In order to do this we rescale HS fields as follows 
\begin{equation}
\omega_{m,n}\longrightarrow\lambda^{-\tfrac{\left|m-n\right|}{2}}\omega_{m,n},\quad C_{m,n}\longrightarrow\lambda^{-\tfrac{m+n}{2}}C_{m,n}.
\end{equation}
For rescaled fields the flat limit $\lambda\rightarrow0$ turn covariant
derivatives to
\begin{eqnarray}
 &  & \mathcal{D}_{ad}\omega\left(Y|K|x\right)\longrightarrow D^{L}\omega+h^{\alpha\dot{\beta}}y_{\alpha}\bar{\partial}_{\dot{\beta}}\omega_{-}+h^{\alpha\dot{\beta}}\partial_{\alpha}\bar{y}_{\dot{\beta}}\omega_{+},\label{D_ad_fl}\\
 &  & \mathcal{D}_{tw}C\left(Y|K|x\right)\longrightarrow D^{L}C+ih^{\alpha\dot{\beta}}\partial_{\alpha}\bar{\partial}_{\dot{\beta}}C,\label{D_tw_fl}
\end{eqnarray}
where $D^{L}$ and $h^{\alpha\dot{\beta}}$ are Lorentz-covariant
derivative and vierbein of Minkowski space-time. Then one substitutes
(\ref{D_ad_fl})-(\ref{D_tw_fl}) into (\ref{On_sh_Th_1})-(\ref{On_sh_Th_2})
and gets linear equations for HS fields in flat space-time.
\begin{eqnarray}
 &  & D^{L}\omega\left(Y|K|x\right)+h^{\alpha\dot{\beta}}y_{\alpha}\bar{\partial}_{\dot{\beta}}\omega_{-}\left(Y|K|x\right)+h^{\alpha\dot{\beta}}\partial_{\alpha}\bar{y}_{\dot{\beta}}\omega_{+}\left(Y|K|x\right)=\dfrac{i}{4}\eta\bar{H}^{\dot{\alpha}\dot{\beta}}\bar{\partial}_{\dot{\alpha}}\bar{\partial}_{\dot{\beta}}C\left(0,\bar{y}|K|x\right)k+\nonumber \\
 &  & +\dfrac{i}{4}\bar{\eta}H^{\alpha\beta}\partial_{\alpha}\partial_{\beta}C\left(y,0|K|x\right)\bar{k},\label{OnShTh_flat_1}\\
 &  & D^{L}C\left(Y|K|x\right)+ih^{\alpha\dot{\beta}}\partial_{\alpha}\bar{\partial}_{\dot{\beta}}C\left(Y|K|x\right)=0.\label{OnShTh_flat_2}
\end{eqnarray}

Now the first step is to express $C$ fields in (\ref{J_s,s}) via
derivatives of Fronsdal fields. To this end one rewrites (\ref{OnShTh_flat_1})
in terms of 0-forms 
\begin{eqnarray}
D^{\beta}\phantom{}_{\dot{\alpha}}\omega_{n,m|\beta\dot{\alpha}} & = & -y^{\beta}\bar{\partial}_{\dot{\alpha}}\left(\omega_{-}\right)_{n-1,m+1|\beta\dot{\alpha}}-\partial^{\beta}\bar{y}_{\dot{\alpha}}\left(\omega_{+}\right)_{n+1,m-1|\beta\dot{\alpha}}+\dfrac{i}{2}\eta\delta_{n,0}\bar{\partial}_{\dot{\alpha}}\bar{\partial}_{\dot{\alpha}}C_{0,m+2}k,\label{On-sh_0_form1}\\
D_{\alpha}\phantom{}^{\dot{\beta}}\omega_{n,m|\alpha\dot{\beta}} & = & -y_{\alpha}\bar{\partial}^{\dot{\beta}}\left(\omega_{-}\right)_{n-1,m+1|\alpha\dot{\beta}}-\partial_{\alpha}\bar{y}^{\dot{\beta}}\left(\omega_{+}\right)_{n+1,m-1|\alpha\dot{\beta}}+\dfrac{i}{2}\bar{\eta}\delta_{m,0}\partial_{\alpha}\partial_{\alpha}C_{n+2,0}\bar{k}.\label{On-sh_0_form2}
\end{eqnarray}
Contracting (\ref{On-sh_0_form1}) with $\bar{y}^{\dot{\alpha}}\bar{y}^{\dot{\alpha}}$
and (\ref{On-sh_0_form2}) with $y^{\alpha}y^{\alpha}$ yields
\begin{eqnarray}
\bar{y}^{\dot{\alpha}}D_{\alpha\dot{\alpha}}\partial^{\alpha}\phi_{n,m} & = & n\cdot m\left(\phi_{-}\right)_{n-1,m+1}-\dfrac{i}{2}\eta\delta_{n,1}m\left(m+1\right)C_{0,m+1}k,\label{fi_rek_1}\\
y^{\alpha}D_{\alpha\dot{\alpha}}\bar{\partial}^{\dot{\alpha}}\phi_{n,m} & = & n\cdot m\left(\phi_{+}\right)_{n+1,m-1}-\dfrac{i}{2}\bar{\eta}\delta_{m,1}n\left(n+1\right)C_{n+1,0}\bar{k}.\label{fi_rek_2}
\end{eqnarray}
From this one finds
\begin{eqnarray}
C_{2s,0} & = & \dfrac{2i\eta}{s\cdot\left(2s\right)!}\left(y^{\alpha}D_{\alpha\dot{\alpha}}\bar{\partial}^{\dot{\alpha}}\right)^{s}\phi_{s,s}\bar{k},\\
C_{0,2s} & = & \dfrac{2i\bar{\eta}}{s\cdot\left(2s\right)!}\left(\bar{y}^{\dot{\alpha}}D_{\alpha\dot{\alpha}}\partial^{\alpha}\right)^{s}\phi_{s,s}k.
\end{eqnarray}
Then (\ref{OnShTh_flat_2}) gives
\begin{eqnarray}
C_{2s+d,d} & = & \dfrac{2\eta\cdot i^{d+1}}{s\cdot\left(2s+d\right)!d!}\left(y^{\beta}D_{\beta\dot{\beta}}\bar{y}^{\dot{\mathbf{\beta}}}\right)^{d}\left(y^{\alpha}D_{\alpha\dot{\alpha}}\bar{\partial}^{\dot{\alpha}}\right)^{s}\phi_{s,s}\bar{k},\label{C_D_fi}\\
C_{d,2s+d} & = & \dfrac{2\bar{\eta}\cdot i^{d+1}}{s\cdot\left(2s+d\right)!d!}\left(y^{\beta}D_{\beta\dot{\beta}}\bar{y}^{\dot{\mathbf{\beta}}}\right)^{d}\left(\bar{y}^{\dot{\alpha}}D_{\alpha\dot{\alpha}}\partial^{\alpha}\right)^{s}\phi_{s,s}k.
\end{eqnarray}
Now one contracts (\ref{unf_1}) with $y^{\alpha}y^{\alpha}$ (or
(\ref{unf_2}) with $\bar{y}^{\dot{\alpha}}\bar{y}^{\dot{\alpha}}$)
and makes use of (\ref{fi_rek_2}) (or (\ref{fi_rek_1})) to obtain
\begin{equation}
\square\phi_{s,s}+...=-s^{2}\left(s-1\right)\mathcal{J}_{s,s}+...,\label{box_eq}
\end{equation}
where $\square=\tfrac{1}{2}D_{\alpha\dot{\alpha}}D^{\alpha\dot{\alpha}}$,
ellipsis on the l.h.s. denotes other terms of Fronsdal kinetic operator
besides the box and ellipsis on the r.h.s. denotes other (gauge-noninvariant)
sources generated by $Q\left(C,\omega\right)$ and $\Gamma_{s<s_{1}+s_{2}}\left(J\right)$
in (\ref{HS_cur_1}).

Now let us consider the current (\ref{J_s,s}). We want to extract
the term describing $s-s_{1}-s_{2}$ vertex. A simple counting shows
that two kind of terms are presented in (\ref{J_s,s}): either two
co-directional helicities are coupled ($C_{+}C_{+}$ or $C_{-}C_{-}$),
then the term has $\left(s+s_{1}+s_{2}\right)$ derivatives, or two
opposite ones ($C_{+}C_{-}$ or $C_{-}C_{+}$), then total number
of derivatives is $\left(s+\left|s_{1}-s_{2}\right|\right)$ (let
us remind that we are in $s\geq s_{1}+s_{2}$ sector). This corresponds
to two types of $4d$ cubic HS vertices found in \cite{Metsaev_2kinds}.
Altogether this means there are no higher-derivative improvements
to vertices of \cite{Metsaev_2kinds}, which could, for instance,
affect locality issue in higher orders. (Note that lower-derivative
improvements to $\left(s+s_{1}+s_{2}\right)$-term cannot contribute
to $\left(s+\left|s_{1}-s_{2}\right|\right)$-term because they have
different helicity structure.) We will analyse two vertices separately.

\subsection{Maximal-derivative part\label{Max_der}}

First, let us consider the part of (\ref{J_s,s}) with $\left(s+s_{1}+s_{2}\right)$
derivatives. This looks as follows
\begin{eqnarray}
 &  & \mathcal{J}_{s-s_{1}-s_{2}}^{H}=i\dfrac{\left(s-2\right)!}{8\left(2s\right)!}\sum_{k,m=0}^{s}\dfrac{\left(m+k\right)!\left(2s-m-k\right)!}{\left(s-k\right)!k!\left(s-m\right)!m!}\left(y^{\alpha}\partial_{\alpha}^{1}\right)^{m}\left(-y^{\beta}\partial_{\beta}^{2}\right)^{s-m}\left(\bar{y}^{\dot{\alpha}}\bar{\partial}_{\dot{\alpha}}^{1}\right)^{s-k}\left(-\bar{y}^{\dot{\beta}}\bar{\partial}_{\dot{\beta}}^{2}\right)^{k}\nonumber \\
 &  & \sum_{n=0}^{s}\dfrac{i^{n}}{\left(s+n-1\right)!}\left(\left(\partial_{\gamma}^{1}\partial^{2\gamma}\right)^{n}+\left(\bar{\partial}_{\dot{\gamma}}^{1}\bar{\partial}^{2\dot{\gamma}}\right)^{n}\right)\sum_{d_{1},d_{2}=0}^{\infty}\biggl\{ C_{2s_{1}+d_{1},d_{1}}\left(Y^{1}|K|x\right)C_{2s_{2}+d_{2},d_{2}}\left(Y^{2}|K|x\right)+\nonumber \\
 &  & +C_{2s_{2}+d_{2},d_{2}}\left(Y^{1}|K|x\right)C_{2s_{1}+d_{1},d_{1}}\left(Y^{2}|K|x\right)+C_{d_{1},2s_{1}+d_{1}}\left(Y^{1}|K|x\right)C_{d_{2},2s_{2}+d_{2}}\left(Y^{2}|K|x\right)+\nonumber \\
 &  & +C_{d_{2},2s_{2}+d_{2}}\left(Y^{1}|K|x\right)C_{d_{1},2s_{1}+d_{1}}\left(Y^{2}|K|x\right)\biggr\}\Bigr|_{Y^{1}=Y^{2}=0}.\label{J_s_s1_s2}
\end{eqnarray}
That spins $s_{1}$, $s_{2}$ of constituent fields are fixed and
all $Y^{1}$ and $Y^{2}$ are eventually put to zero reduces the fivefold
sum in (\ref{J_s_s1_s2}) to the single one:
\begin{eqnarray}
\mathcal{J}_{s-s_{1}-s_{2}}^{H} & = & i\dfrac{\left(s-2\right)!}{8\left(2s\right)!}\sum_{d=0}^{s}\dfrac{\left(s+s_{1}-s_{2}\right)!\left(s-s_{1}+s_{2}\right)!}{\left(s-d\right)!d!\left(-s_{1}+s_{2}+d\right)!\left(s+s_{1}-s_{2}-d\right)!}\nonumber \\
 &  & \left(y^{\alpha}\partial_{\alpha}^{1}\right)^{s+s_{1}-s_{2}-d}\left(y^{\beta}\partial_{\beta}^{2}\right)^{-s_{1}+s_{2}+d}\left(\bar{y}^{\dot{\alpha}}\bar{\partial}_{\dot{\alpha}}^{1}\right)^{s-d}\left(\bar{y}^{\dot{\beta}}\bar{\partial}_{\dot{\beta}}^{2}\right)^{d}\dfrac{i^{s_{1}+s_{2}}\left(-1\right)^{s+d}\left(1+\left(-1\right)^{s+s_{1}+s_{2}}\right)}{\left(s+s_{1}+s_{2}-1\right)!}\nonumber \\
 &  & \left(\left(\partial_{\gamma}^{1}\partial^{2\gamma}\right)^{s_{1}+s_{2}}C_{2s_{1}+s-d,s-d}\left(Y^{1}|x\right)C_{2s_{2}+d,d}\left(Y^{2}|x\right)+h.c.\right)\Bigr|_{Y^{1}=Y^{2}=0}
\end{eqnarray}
(here we resolved $K$-dependence using (\ref{extKlein})), that after
evaluating derivatives from the second line yields
\begin{eqnarray}
\mathcal{J}_{s-s_{1}-s_{2}}^{H} & = & \dfrac{\left(s-2\right)!\left(s+s_{1}-s_{2}\right)!\left(s-s_{1}+s_{2}\right)!}{8\left(2s\right)!\left(s+s_{1}+s_{2}-1\right)!}i^{s_{1}+s_{2}+1}\left(1+\left(-1\right)^{s+s_{1}+s_{2}}\right)\cdot\nonumber \\
 &  & \cdot\sum_{d=0}^{s}\left(-1\right)^{s+d}\biggl\{\left(\partial_{\gamma}\right)^{s_{1}+s_{2}}C_{2s_{1}+s-d,s-d}\left(Y|x\right)\cdot\left(\partial^{\gamma}\right)^{s_{1}+s_{2}}C_{2s_{2}+d,d}\left(Y|x\right)+\nonumber \\
 &  & +\left(\bar{\partial}_{\dot{\gamma}}\right)^{s_{1}+s_{2}}C_{2s_{1}+s-d,s-d}\left(Y|x\right)\cdot\left(\bar{\partial}^{\dot{\gamma}}\right)^{s_{1}+s_{2}}C_{2s_{2}+d,d}\left(Y|x\right)\biggr\}.\label{J_s,s-1}
\end{eqnarray}
Note that due to $\left(1+\left(-1\right)^{s+s_{1}+s_{2}}\right)$
factor, (\ref{J_s,s-1}) vanishes if the total sum of spins is odd.
In fact, this is because we have only one field of every spin, similarly
to the electrodynamics where one needs two copies of the matter fields
to have a nonzero electric current. So if one considers matrix-valued
HS fields, the contribution would be nonzero.

Now let us analyse the spinorial expression in (\ref{J_s,s-1}). Our
goal is to bring it to the form that can be simply re-expressed in
terms of Lorentz tensors.

First, we use (\ref{C_D_fi}) to rewrite it as
\begin{eqnarray}
 &  & \left(\partial_{\gamma}\right)^{s_{1}+s_{2}}C_{2s_{1}+s-d,s-d}\left(Y|x\right)\cdot\left(\partial^{\gamma}\right)^{s_{1}+s_{2}}C_{2s_{2}+d,d}\left(Y|x\right)=-\dfrac{4\eta^{2}i^{s}}{\left(2s_{1}+s-d\right)!\left(2s_{2}+d\right)!\left(s-d\right)!d!\cdot s_{1}\cdot s_{2}}\nonumber \\
 &  & \left\{ \left(\partial_{\gamma}\right)^{s_{1}+s_{2}}\left(y^{\alpha}D_{\alpha\dot{\alpha}}\bar{y}^{\dot{\mathbf{\alpha}}}\right)^{s-d}\left(y^{\alpha}D_{\alpha\dot{\alpha}}\bar{\partial}^{\dot{\alpha}}\right)^{s_{1}}\phi_{s_{1},s_{1}}\right\} \left\{ \left(\partial^{\gamma}\right)^{s_{1}+s_{2}}\left(y^{\beta}D_{\beta\dot{\beta}}\bar{y}^{\dot{\mathbf{\beta}}}\right)^{d}\left(y^{\beta}D_{\beta\dot{\beta}}\bar{\partial}^{\dot{\beta}}\right)^{s_{2}}\phi_{s_{2},s_{2}}\right\} .\label{dC_dC}
\end{eqnarray}
Evaluating spinorial derivatives gives
\begin{eqnarray}
 &  & \left\{ \left(\partial_{\gamma}\right)^{s_{1}+s_{2}}\left(y^{\alpha}D_{\alpha\dot{\alpha}}\bar{y}^{\dot{\mathbf{\alpha}}}\right)^{s-d}\left(y^{\alpha}D_{\alpha\dot{\alpha}}\bar{\partial}^{\dot{\alpha}}\right)^{s_{1}}\phi_{s_{1},s_{1}}\right\} \left\{ \left(\partial^{\gamma}\right)^{s_{1}+s_{2}}\left(y^{\beta}D_{\beta\dot{\beta}}\bar{y}^{\dot{\mathbf{\beta}}}\right)^{d}\left(y^{\beta}D_{\beta\dot{\beta}}\bar{\partial}^{\dot{\beta}}\right)^{s_{2}}\phi_{s_{2},s_{2}}\right\} =\nonumber \\
 &  & =\dfrac{s_{1}!s_{2}!\left(s+2s_{1}-d\right)!\left(2s_{2}+d\right)!}{\left(s+s_{1}-s_{2}-d\right)!\left(-s_{1}+s_{2}+d\right)!}\left\{ \left(\delta_{\gamma}\phantom{}^{\mu}\right)^{s_{1}+s_{2}}\left(y^{\mu}\right)^{s+s_{1}-s_{2}-d}\left(\bar{y}^{\dot{\mu}}\right)^{s-d}\left(D_{\mu\dot{\mu}}\right)^{s-d}\left(D_{\mu\dot{\beta}}\right)^{s_{1}}\phi_{\mu\left(s_{1}\right),}\phantom{}^{\dot{\beta}\left(s_{1}\right)}\right\} \nonumber \\
 &  & \left\{ \left(\epsilon^{\gamma\nu}\right)^{s_{1}+s_{2}}\left(y^{\nu}\right)^{-s_{1}+s_{2}+d}\left(\bar{y}^{\dot{\nu}}\right)^{d}\left(D_{\nu\dot{\nu}}\right)^{d}\left(D_{\nu\dot{\alpha}}\right)^{s_{2}}\phi_{\nu\left(s_{2}\right),}\phantom{}^{\dot{\alpha}\left(s_{2}\right)}\right\} .\label{J_intm_1}
\end{eqnarray}
Due to symmetrisation over $\mu$ (and over $\nu$), $\gamma$ indices
after applying $\left(\delta_{\gamma}\phantom{}^{\mu}\right)^{s_{1}+s_{2}}$
and $\left(\epsilon^{\gamma\nu}\right)^{s_{1}+s_{2}}$ will hang symmetrically
on fields and derivatives. But we are going to arrange gammas in some
particular order. To this end we establish some useful relations.
The first is (we write down only relevant indices)
\begin{equation}
D_{\alpha\dot{\gamma}}\phi_{\beta}\phantom{}^{\dot{\gamma}}=D_{\beta\dot{\gamma}}\phi_{\alpha}\phantom{}^{\dot{\gamma}}+\epsilon_{\alpha\beta}D_{\gamma\dot{\gamma}}\phi^{\gamma\dot{\gamma}}\approx D_{\beta\dot{\gamma}}\phi_{\alpha}\phantom{}^{\dot{\gamma}},\label{id1}
\end{equation}
where the approximate equality symbol means that we lopped off a divergence
of the field, as we neglect it in our problem. The second is
\begin{equation}
D_{\alpha\dot{\alpha}}D_{\beta\dot{\beta}}\phi\approx D_{\alpha\dot{\beta}}D_{\beta\dot{\alpha}}\phi,\label{id2}
\end{equation}
which is modulo boxes (that can be redefined away) and terms with
$D_{\alpha\dot{\beta}}D_{\alpha}\phantom{}^{\mathbf{\dot{\beta}}}$
($D_{\beta\dot{\alpha}}D^{\beta}\phantom{}_{\dot{\alpha}}$), which
are zeros in flat space. Using these two relations, we obtain the
third one
\begin{equation}
D_{\gamma\dot{\gamma}}D_{\alpha\dot{\beta}}\phi_{\beta}\phantom{}^{\dot{\beta}}\approx D_{\beta\dot{\gamma}}D_{\alpha\dot{\beta}}\phi_{\gamma}\phantom{}^{\dot{\beta}}.\label{id3}
\end{equation}
Altogether they imply that we are free to put gammas on any places
instead of lower $\mu$ ($\nu$) indices in (\ref{J_intm_1}) as all
combinations are equivalent. So, assuming for definiteness $s_{1}\geq s_{2}$,
we rewrite (\ref{J_intm_1}) as
\begin{eqnarray}
 &  & \dfrac{s_{1}!s_{2}!\left(s+2s_{1}-d\right)!\left(2s_{2}+d\right)!}{\left(s+s_{1}-s_{2}-d\right)!\left(-s_{1}+s_{2}+d\right)!}\left(y^{\mu}\right)^{s}\left(\bar{y}^{\dot{\mu}}\right)^{s}\left\{ \left(D_{\mu\dot{\mu}}\right)^{s-d}\left(D_{\mu\dot{\beta}}\right)^{s_{1}-s_{2}}\left(D_{\gamma\dot{\beta}}\right)^{s_{2}}\phi_{\gamma\left(s_{1}\right),}\phantom{}^{\dot{\beta}\left(s_{1}\right)}\right\} \nonumber \\
 &  & \left\{ \left(D_{\mu\dot{\mu}}\right)^{d-s_{1}+s_{2}}\left(D^{\gamma}\phantom{}_{\dot{\mu}}\right)^{s_{1}-s_{2}}\left(D^{\gamma}\phantom{}_{\dot{\alpha}}\right)^{s_{2}}\phi^{\gamma\left(s_{2}\right),\dot{\alpha}\left(s_{2}\right)}\right\} ,
\end{eqnarray}
and, using (\ref{id1}), further as
\begin{eqnarray}
 &  & \dfrac{s_{1}!s_{2}!\left(s+2s_{1}-d\right)!\left(2s_{2}+d\right)!}{\left(s+s_{1}-s_{2}-d\right)!\left(-s_{1}+s_{2}+d\right)!}\left(y^{\mu}\right)^{s}\left(\bar{y}^{\dot{\mu}}\right)^{s}\left\{ \left(D_{\mu\dot{\mu}}\right)^{s-d}\left(D_{\mu\dot{\beta}}\right)^{s_{1}-s_{2}}\left(D_{\delta\dot{\beta}}\right)^{s_{2}}\phi_{\gamma\left(s_{1}\right),}\phantom{}^{\dot{\beta}\left(s_{1}\right)}\right\} \nonumber \\
 &  & \left\{ \left(D_{\mu\dot{\mu}}\right)^{d-s_{1}+s_{2}}\left(D^{\gamma}\phantom{}_{\dot{\mu}}\right)^{s_{1}-s_{2}}\left(D^{\gamma}\phantom{}_{\dot{\alpha}}\right)^{s_{2}}\phi^{\delta\left(s_{2}\right),\dot{\alpha}\left(s_{2}\right)}\right\} .\label{J_intm_2}
\end{eqnarray}
Now we want to replace all lower $\dot{\beta}$ in $\left(D_{\delta\dot{\beta}}\right)^{s_{2}}$
in the first line with lower $\dot{\alpha}$ so as to make these derivatives
to be entirely contracted with the spin-$s_{2}$ field. We can perform
this with the help of $D_{\mu\dot{\beta}}$ in the first line, because
\begin{eqnarray}
 &  & D_{\mu\dot{\beta}}D_{\delta\dot{\beta}}\phi^{\gamma\dot{\beta}\dot{\beta}}\cdot D_{\gamma\dot{\alpha}}\phi^{\delta\dot{\alpha}}=D_{\mu\dot{\beta}}D_{\delta\dot{\alpha}}\phi^{\gamma\dot{\beta}\dot{\beta}}\cdot D_{\gamma\dot{\beta}}\phi^{\delta\dot{\alpha}}+D_{\mu\dot{\beta}}D_{\delta\dot{\gamma}}\phi^{\gamma\dot{\beta}}\phantom{}_{\dot{\alpha}}\cdot D_{\gamma}\phantom{}^{\dot{\gamma}}\phi^{\delta\dot{\alpha}}\approx\nonumber \\
 &  & \approx D_{\mu\dot{\beta}}D_{\delta\dot{\alpha}}\phi^{\gamma\dot{\beta}\dot{\beta}}\cdot D_{\gamma\dot{\beta}}\phi^{\delta\dot{\alpha}}+D_{\mu\dot{\beta}}D_{\gamma\dot{\gamma}}\phi^{\gamma\dot{\beta}}\phantom{}_{\dot{\alpha}}\cdot D_{\delta}\phantom{}^{\dot{\gamma}}\phi^{\delta\dot{\alpha}}\approx D_{\mu\dot{\beta}}D_{\delta\dot{\alpha}}\phi^{\gamma\dot{\beta}\dot{\beta}}\cdot D_{\gamma\dot{\beta}}\phi^{\delta\dot{\alpha}},\label{rel1}
\end{eqnarray}
where at the penultimate step we used that
\begin{eqnarray}
 &  & D_{\alpha\dot{\gamma}}\phi\cdot D_{\beta}\phantom{}^{\dot{\gamma}}\phi=D_{\beta\dot{\gamma}}\phi\cdot D_{\alpha}\phantom{}^{\dot{\gamma}}\phi+\epsilon_{\alpha\beta}D_{\gamma\dot{\gamma}}\phi\cdot D^{\gamma\dot{\gamma}}\phi=\nonumber \\
 &  & =D_{\beta\dot{\gamma}}\phi\cdot D_{\alpha}\phantom{}^{\dot{\gamma}}\phi+\epsilon_{\alpha\beta}\square\left(\phi\cdot\phi\right)-\epsilon_{\alpha\beta}\left(\square\phi\cdot\phi+\phi\cdot\square\phi\right)\approx\nonumber \\
 &  & \approx D_{\beta\dot{\gamma}}\phi\cdot D_{\alpha}\phantom{}^{\dot{\gamma}}\phi,\label{id4}
\end{eqnarray}
and at the last step that
\begin{equation}
D_{\alpha\dot{\alpha}}D_{\beta\dot{\beta}}\phi^{\alpha\dot{\beta}}\approx D_{\beta\dot{\alpha}}D_{\alpha\dot{\beta}}\phi^{\alpha\dot{\beta}}\approx0.\label{id5}
\end{equation}
Thus (\ref{J_intm_2}) turns into
\begin{eqnarray}
 &  & \dfrac{s_{1}!s_{2}!\left(s+2s_{1}-d\right)!\left(2s_{2}+d\right)!}{\left(s+s_{1}-s_{2}-d\right)!\left(-s_{1}+s_{2}+d\right)!}\left(y^{\mu}\right)^{s}\left(\bar{y}^{\dot{\mu}}\right)^{s}\left\{ \left(D_{\mu\dot{\mu}}\right)^{s-d}\left(D_{\mu\dot{\beta}}\right)^{s_{1}-s_{2}}\left(D_{\delta\dot{\alpha}}\right)^{s_{2}}\phi_{\gamma\left(s_{1}\right),}\phantom{}^{\dot{\beta}\left(s_{1}\right)}\right\} \nonumber \\
 &  & \left\{ \left(D_{\mu\dot{\mu}}\right)^{d-s_{1}+s_{2}}\left(D^{\gamma}\phantom{}_{\dot{\mu}}\right)^{s_{1}-s_{2}}\left(D^{\gamma}\phantom{}_{\dot{\beta}}\right)^{s_{2}}\phi^{\delta\left(s_{2}\right),\dot{\alpha}\left(s_{2}\right)}\right\} .
\end{eqnarray}
Now we want to exchange $\dot{\beta}$ in $\left(D_{\mu\dot{\beta}}\right)^{s_{1}-s_{2}}$
in the first line with $\dot{\mu}$ in $\left(D^{\gamma}\phantom{}_{\dot{\mu}}\right)^{s_{1}-s_{2}}$
in the second line. This can be done by virtue of a relation, similar
to (\ref{rel1}):
\begin{eqnarray}
 &  & D_{\mu\dot{\beta}}D_{\mu\dot{\beta}}\phi^{\gamma\dot{\beta}\dot{\beta}}\cdot D_{\gamma\dot{\mu}}\phi=D_{\mu\dot{\beta}}D_{\mu\dot{\mu}}\phi^{\gamma\dot{\beta}\dot{\beta}}\cdot D_{\gamma\dot{\beta}}\phi+D_{\mu\dot{\beta}}D_{\mu\dot{\alpha}}\phi^{\gamma\dot{\beta}}\phantom{}_{\dot{\mu}}\cdot D_{\gamma}\phantom{}^{\dot{\alpha}}\phi\approx\nonumber \\
 &  & \approx D_{\mu\dot{\beta}}D_{\mu\dot{\mu}}\phi^{\gamma\dot{\beta}\dot{\beta}}\cdot D_{\gamma\dot{\beta}}\phi+D_{\mu\dot{\beta}}D_{\gamma\dot{\alpha}}\phi^{\gamma\dot{\beta}}\phantom{}_{\dot{\mu}}\cdot D_{\mu}\phantom{}^{\dot{\alpha}}\phi\approx D_{\mu\dot{\beta}}D_{\mu\dot{\mu}}\phi^{\gamma\dot{\beta}\dot{\beta}}\cdot D_{\gamma\dot{\beta}}\phi.\label{id6}
\end{eqnarray}
This allows us to perform all necessary exchanges in $\left(D_{\mu\dot{\beta}}\right)^{s_{1}-s_{2}}$
except for the last one, because to use (\ref{id6}) we need at least
two $D_{\mu\dot{\beta}}$. So we have
\begin{eqnarray}
 &  & \dfrac{s_{1}!s_{2}!\left(s+2s_{1}-d\right)!\left(2s_{2}+d\right)!}{\left(s+s_{1}-s_{2}-d\right)!\left(-s_{1}+s_{2}+d\right)!}\left(y^{\mu}\right)^{s}\left(\bar{y}^{\dot{\mu}}\right)^{s}\left\{ D_{\mu\dot{\beta}}\left(D_{\mu\dot{\mu}}\right)^{s+s_{1}-s_{2}-d-1}\left(D_{\alpha\dot{\alpha}}\right)^{s_{2}}\phi_{\beta\left(s_{1}\right),}\phantom{}^{\dot{\beta}\left(s_{1}\right)}\right\} \nonumber \\
 &  & \left\{ D^{\beta}\phantom{}_{\dot{\mu}}\left(D_{\mu\dot{\mu}}\right)^{-s_{1}+s_{2}+d}\left(D^{\beta}\phantom{}_{\dot{\beta}}\right)^{s_{1}-1}\phi^{\alpha\left(s_{2}\right),\dot{\alpha}\left(s_{2}\right)}\right\} ,\label{J_intm_3}
\end{eqnarray}
and the last exchange leads to the expression of the form
\begin{eqnarray}
 &  & \dfrac{s_{1}!s_{2}!\left(s+2s_{1}-d\right)!\left(2s_{2}+d\right)!}{\left(s+s_{1}-s_{2}-d\right)!\left(-s_{1}+s_{2}+d\right)!}\left(y^{\mu}\right)^{s}\left(\bar{y}^{\dot{\mu}}\right)^{s}\left(-1\right)^{s_{1}}\nonumber \\
 &  & \biggl[\left(D_{\mu\dot{\mu}}\right)^{s+s_{1}-s_{2}-d}\left(D_{\alpha\dot{\alpha}}\right)^{s_{2}}\phi^{\beta\left(s_{1}\right),\dot{\beta}\left(s_{1}\right)}\cdot\left(D_{\mu\dot{\mu}}\right)^{-s_{1}+s_{2}+d}\left(D_{\beta\dot{\beta}}\right)^{s_{1}}\phi^{\alpha\left(s_{2}\right),\dot{\alpha}\left(s_{2}\right)}-\nonumber \\
 &  & -\left(D_{\mu\dot{\mu}}\right)^{s+s_{1}-s_{2}-d-1}\left(D_{\alpha\dot{\alpha}}\right)^{s_{2}}D_{\beta\dot{\mu}}\phi^{\beta\left(s_{1}\right),\dot{\beta}\left(s_{1}\right)}\cdot D_{\mu\dot{\beta}}\left(D_{\mu\dot{\mu}}\right)^{-s_{1}+s_{2}+d}\left(D_{\beta\dot{\beta}}\right)^{s_{1}-1}\phi^{\alpha\left(s_{2}\right),\dot{\alpha}\left(s_{2}\right)}\biggr].\label{J_intm_4}
\end{eqnarray}

Now, substituting (\ref{J_intm_4}) for the second line in (\ref{dC_dC}),
then (\ref{dC_dC}) in (\ref{J_s,s-1}), adding conjugate expression
and simplifying, one gets
\begin{eqnarray}
 &  & \mathcal{J}_{s-s_{1}-s_{2}}^{H}=-i\dfrac{\left(s-2\right)!}{4\left(2s\right)!}\sum_{d=0}^{s}\left(\begin{array}{c}
s+s_{1}-s_{2}\\
d
\end{array}\right)\left(\begin{array}{c}
s-s_{1}+s_{2}\\
s-d
\end{array}\right)\left(s_{1}-1\right)!\left(s_{2}-1\right)!\nonumber \\
 &  & \dfrac{i^{s+s_{1}+s_{2}}\left(-1\right)^{s+d+s_{1}}\left(1+\left(-1\right)^{s+s_{1}+s_{2}}\right)}{\left(s+s_{1}+s_{2}-1\right)!}\left(y^{\mu}\right)^{s}\left(\bar{y}^{\dot{\mu}}\right)^{s}\nonumber \\
 &  & \biggl\{\left(\eta^{2}+\bar{\eta}^{2}\right)\left(D_{\mu\dot{\mu}}\right)^{s+s_{1}-s_{2}-d}\left(D_{\alpha\dot{\alpha}}\right)^{s_{2}}\phi^{\beta\left(s_{1}\right),\dot{\beta}\left(s_{1}\right)}\cdot\left(D_{\mu\dot{\mu}}\right)^{-s_{1}+s_{2}+d}\left(D_{\beta\dot{\beta}}\right)^{s_{1}}\phi^{\alpha\left(s_{2}\right),\dot{\alpha}\left(s_{2}\right)}+\nonumber \\
 &  & +\left(\eta^{2}-\bar{\eta}^{2}\right)\biggl[D_{\mu\dot{\beta}}\left(D_{\mu\dot{\mu}}\right)^{s+s_{1}-s_{2}-d-1}\left(D_{\alpha\dot{\alpha}}\right)^{s_{2}}\phi^{\beta\left(s_{1}\right),\dot{\beta}\left(s_{1}\right)}\cdot D_{\beta\dot{\mu}}\left(D_{\mu\dot{\mu}}\right)^{-s_{1}+s_{2}+d}\left(D_{\beta\dot{\beta}}\right)^{s_{1}-1}\phi^{\alpha\left(s_{2}\right),\dot{\alpha}\left(s_{2}\right)}-\nonumber \\
 &  & -D_{\beta\dot{\mu}}\left(D_{\mu\dot{\mu}}\right)^{s+s_{1}-s_{2}-d-1}\left(D_{\alpha\dot{\alpha}}\right)^{s_{2}}\phi^{\beta\left(s_{1}\right),\dot{\beta}\left(s_{1}\right)}\cdot D_{\mu\dot{\beta}}\left(D_{\mu\dot{\mu}}\right)^{-s_{1}+s_{2}+d}\left(D_{\beta\dot{\beta}}\right)^{s_{1}-1}\phi^{\alpha\left(s_{2}\right),\dot{\alpha}\left(s_{2}\right)}\biggr]\biggr\}.\label{J_spin}
\end{eqnarray}
Then, using Vandermonde's identity 
\begin{equation}
\sum_{n=0}^{c}\left(\begin{array}{c}
a\\
n
\end{array}\right)\left(\begin{array}{c}
b\\
c-n
\end{array}\right)=\left(\begin{array}{c}
a+b\\
c
\end{array}\right)\label{Vdm}
\end{equation}
and integrating by parts one can perform a summation over $d$ explicitly,
reducing (\ref{J_spin}) to
\begin{eqnarray}
 &  & \mathcal{J}_{s-s_{1}-s_{2}}^{H}=-\dfrac{i^{s+s_{1}+s_{2}+1}\left(s-2\right)!\left(s_{1}-1\right)!\left(s_{2}-1\right)!\left(-1\right)^{s_{1}}\left(1+\left(-1\right)^{s+s_{1}+s_{2}}\right)}{4\cdot s!s!\left(s+s_{1}+s_{2}-1\right)!}\left(y^{\mu}\right)^{s}\left(\bar{y}^{\dot{\mu}}\right)^{s}\nonumber \\
 &  & \biggl\{\left(\eta^{2}+\bar{\eta}^{2}\right)\left(D_{\mu\dot{\mu}}\right)^{s}\left(D_{\alpha\dot{\alpha}}\right)^{s_{2}}\phi^{\beta\left(s_{1}\right),\dot{\beta}\left(s_{1}\right)}\cdot\left(D_{\beta\dot{\beta}}\right)^{s_{1}}\phi^{\alpha\left(s_{2}\right),\dot{\alpha}\left(s_{2}\right)}+\nonumber \\
 &  & +\left(\eta^{2}-\bar{\eta}^{2}\right)\biggl[D_{\mu\dot{\beta}}\left(D_{\mu\dot{\mu}}\right)^{s-1}\left(D_{\alpha\dot{\alpha}}\right)^{s_{2}}\phi^{\beta\left(s_{1}\right),\dot{\beta}\left(s_{1}\right)}\cdot D_{\beta\dot{\mu}}\left(D_{\beta\dot{\beta}}\right)^{s_{1}-1}\phi^{\alpha\left(s_{2}\right),\dot{\alpha}\left(s_{2}\right)}-\nonumber \\
 &  & -D_{\beta\dot{\mu}}\left(D_{\mu\dot{\mu}}\right)^{s-1}\left(D_{\alpha\dot{\alpha}}\right)^{s_{2}}\phi^{\beta\left(s_{1}\right),\dot{\beta}\left(s_{1}\right)}\cdot D_{\mu\dot{\beta}}\left(D_{\beta\dot{\beta}}\right)^{s_{1}-1}\phi^{\alpha\left(s_{2}\right),\dot{\alpha}\left(s_{2}\right)}\biggr]\biggr\}.\label{J_H_fin}
\end{eqnarray}
Here we reached our goal because this expression can easily be translated
into Lorentz tensors as we show below. Now we are going to process
another part of the current (\ref{J_s,s}) that contains $\left(s+\left|s_{1}-s_{2}\right|\right)$
derivatives.

\subsection{Minimal-derivative part}

Analysis of the minimal-derivative part of (\ref{J_s,s}) which has
the form
\begin{eqnarray}
 &  & \mathcal{J}_{s-s_{1}-s_{2}}^{L}=i\dfrac{\left(s-2\right)!}{8\left(2s\right)!}\sum_{k,m=0}^{s}\dfrac{\left(m+k\right)!\left(2s-m-k\right)!}{\left(s-k\right)!k!\left(s-m\right)!m!}\left(y^{\alpha}\partial_{\alpha}^{1}\right)^{m}\left(-y^{\beta}\partial_{\beta}^{2}\right)^{s-m}\left(\bar{y}^{\dot{\alpha}}\bar{\partial}_{\dot{\alpha}}^{1}\right)^{s-k}\left(-\bar{y}^{\dot{\beta}}\bar{\partial}_{\dot{\beta}}^{2}\right)^{k}\nonumber \\
 &  & \sum_{n=0}^{s}\dfrac{i^{n}}{\left(s+n-1\right)!}\left(\left(\partial_{\gamma}^{1}\partial^{2\gamma}\right)^{n}+\left(\bar{\partial}_{\dot{\gamma}}^{1}\bar{\partial}^{2\dot{\gamma}}\right)^{n}\right)\sum_{d_{1},d_{2}=0}^{\infty}\biggl\{ C_{2s_{1}+d_{1},d_{1}}\left(Y^{1}|K|x\right)C_{d_{2},2s_{2}+d_{2}}\left(Y^{2}|K|x\right)+\nonumber \\
 &  & +C_{2s_{2}+d_{2},d_{2}}\left(Y^{1}|K|x\right)C_{d_{1},2s_{1}+d_{1}}\left(Y^{2}|K|x\right)+C_{d_{1},2s_{1}+d_{1}}\left(Y^{1}|K|x\right)C_{2s_{2}+d_{2},d_{2}}\left(Y^{2}|K|x\right)+\nonumber \\
 &  & +C_{d_{2},2s_{2}+d_{2}}\left(Y^{1}|K|x\right)C_{2s_{1}+d_{1},d_{1}}\left(Y^{2}|K|x\right)\biggr\}\Bigr|_{Y^{1}=Y^{2}=0},\label{JL_s_s1_s2}
\end{eqnarray}
practically repeats analysis of the maximal-derivative one.

First, one evaluates derivatives from the first line of (\ref{JL_s_s1_s2})
and simplifies the expression to
\begin{eqnarray}
\mathcal{J}_{s-s_{1}-s_{2}}^{L} & = & \dfrac{\left(s-2\right)!\left(s+s_{1}+s_{2}\right)!\left(s-s_{1}-s_{2}\right)!}{8\cdot\left(2s\right)!\left(s+s_{1}-s_{2}-1\right)!}i^{s_{1}+s_{2}+1}\left(1+\left(-1\right)^{s+s_{1}+s_{2}}\right)\cdot\nonumber \\
 &  & \sum_{d=0}^{s}\left(-1\right)^{s+d+s_{1}}\biggl\{\left(\partial_{\gamma}\right)^{s_{1}-s_{2}}C_{2s_{1}+d,d}\left(Y|x\right)\cdot\left(\partial^{\gamma}\right)^{s_{1}-s_{2}}C_{s-2s_{2}-d,s-d}\left(Y|x\right)+\nonumber \\
 &  & +\left(\bar{\partial}_{\dot{\gamma}}\right)^{s_{1}-s_{2}}C_{d,2s_{1}+d}\left(Y|x\right)\cdot\left(\bar{\partial}^{\dot{\gamma}}\right)^{s_{1}-s_{2}}C_{s-d,s-2s_{2}-d}\left(Y|x\right)\biggr\}.\label{JL_s,s-1}
\end{eqnarray}
Then, using (\ref{C_D_fi}), one rewrites the first term in brackets
in (\ref{JL_s,s-1}) as
\begin{eqnarray}
 &  & \left(\partial_{\gamma}\right)^{s_{1}-s_{2}}C_{2s_{1}+d,d}\left(Y|x\right)\cdot\left(\partial^{\gamma}\right)^{s_{1}-s_{2}}C_{s-2s_{2}-d,s-d}\left(Y|x\right)=-\dfrac{4i^{s}\left(-1\right)^{s_{2}}\left(s_{1}-1\right)!\left(s_{2}-1\right)!}{\left(s_{1}+s_{2}+d\right)!\left(s-s_{1}-s_{2}-d\right)!\left(s-d\right)!d!}\cdot\nonumber \\
 &  & \cdot\left\{ \left(\delta_{\gamma}\phantom{}^{\mu}\right)^{s_{1}-s_{2}}\left(y^{\mu}\right)^{s_{1}+s_{2}+d}\left(\bar{y}^{\dot{\mu}}\right)^{d}\left(D_{\mu\dot{\mu}}\right)^{d}\left(D_{\mu\dot{\alpha}}\right)^{s_{1}}\phi_{\mu\left(s_{1}\right),}\phantom{}^{\dot{\alpha}\left(s_{1}\right)}\right\} \cdot\nonumber \\
 &  & \cdot\left\{ \left(\epsilon^{\gamma\nu}\right)^{s_{1}-s_{2}}\left(y^{\nu}\right)^{s-s_{1}-s_{2}-d}\left(\bar{y}^{\dot{\nu}}\right)^{s-d}\left(D_{\nu\dot{\nu}}\right)^{s-2s_{2}-d}\left(D_{\alpha\dot{\nu}}\right)^{s_{2}}\phi^{\alpha\left(s_{2}\right)}\phantom{}_{\dot{\nu}\left(s_{2}\right)}\right\} .\label{dC_dC_L}
\end{eqnarray}
As in Section \ref{Max_der}, by means of (\ref{id1})-(\ref{id3})
one hangs all gammas in the second line of (\ref{dC_dC_L}) on spin-$s_{1}$
field
\begin{eqnarray}
 &  & \left(\partial_{\gamma}\right)^{s_{1}-s_{2}}C_{2s_{1}+d,d}\left(Y|x\right)\cdot\left(\partial^{\gamma}\right)^{s_{1}-s_{2}}C_{s-2s_{2}-d,s-d}\left(Y|x\right)=-\dfrac{4i^{s}\left(-1\right)^{s_{1}}\left(s_{1}-1\right)!\left(s_{2}-1\right)!}{\left(s_{1}+s_{2}+d\right)!\left(s-s_{1}-s_{2}-d\right)!\left(s-d\right)!d!}\cdot\nonumber \\
 &  & \cdot\left(y^{\mu}\right)^{s}\left(\bar{y}^{\dot{\mu}}\right)^{s}\left\{ \left(D_{\mu\dot{\mu}}\right)^{d}\left(D_{\mu\dot{\beta}}\right)^{s_{1}}\phi_{\mu\left(s_{2}\right)}\phantom{}^{\gamma\left(s_{1}-s_{2}\right),\dot{\beta}\left(s_{1}\right)}\right\} \left\{ \left(D_{\gamma\dot{\mu}}\right)^{s_{1}-s_{2}}\left(D_{\mu\dot{\mu}}\right)^{s-s_{1}-s_{2}-d}\left(D_{\alpha\dot{\mu}}\right)^{s_{2}}\phi^{\alpha\left(s_{2}\right),}\phantom{}_{\dot{\mu}\left(s_{2}\right)}\right\} ,\nonumber \\
\label{J_intm_1-1}
\end{eqnarray}
and exchanges $(s_{1}-s_{2})$ pieces of $\dot{\beta}$ of $D_{\mu\dot{\beta}}$
in the first bracket with $\dot{\mu}$ of $D_{\gamma\dot{\mu}}$ from
the second bracket 
\begin{eqnarray}
 &  & \left\{ \left(D_{\mu\dot{\mu}}\right)^{d}\left(D_{\mu\dot{\beta}}\right)^{s_{1}}\phi_{\mu\left(s_{2}\right)}\phantom{}^{\gamma\left(s_{1}-s_{2}\right),\dot{\beta}\left(s_{1}\right)}\right\} \left\{ \left(D_{\gamma\dot{\mu}}\right)^{s_{1}-s_{2}}\left(D_{\mu\dot{\mu}}\right)^{s-s_{1}-s_{2}-d}\left(D_{\alpha\dot{\mu}}\right)^{s_{2}}\phi^{\alpha\left(s_{2}\right),}\phantom{}_{\dot{\mu}\left(s_{2}\right)}\right\} =\nonumber \\
 &  & =\left\{ \left(D_{\mu\dot{\mu}}\right)^{s_{1}-s_{2}+d}\left(D_{\mu\dot{\beta}}\right)^{s_{2}}\phi_{\mu\left(s_{2}\right)}\phantom{}^{\gamma\left(s_{1}-s_{2}\right),\dot{\beta}\left(s_{1}\right)}\right\} \left\{ \left(D_{\gamma\dot{\beta}}\right)^{s_{1}-s_{2}}\left(D_{\mu\dot{\mu}}\right)^{s-s_{1}-s_{2}-d}\left(D_{\alpha\dot{\mu}}\right)^{s_{2}}\phi^{\alpha\left(s_{2}\right),}\phantom{}_{\dot{\mu}\left(s_{2}\right)}\right\} .\nonumber \\
\end{eqnarray}
Substituting all this into (\ref{JL_s,s-1}), adding conjugate term
and allowing for
\begin{equation}
D_{\mu\dot{\beta}}\phi_{\mu}\phantom{}^{\dot{\beta}}\cdot D_{\alpha\dot{\mu}}\phi^{\alpha}\phantom{}_{\dot{\mu}}\approx D_{\mu\dot{\mu}}\phi^{\alpha\dot{\beta}}\cdot D_{\alpha\dot{\beta}}\phi_{\mu\dot{\mu}}-D_{\mu\dot{\mu}}\phi^{\alpha\dot{\beta}}\cdot D_{\mu\dot{\mu}}\phi_{\alpha\dot{\beta}}-D_{\beta\dot{\mu}}\phi^{\beta}\phantom{}_{\dot{\mu}}\cdot D_{\mu\dot{\alpha}}\phi_{\mu}\phantom{}^{\dot{\alpha}}\label{rel1-1}
\end{equation}
leads to the following expression for the minimal-derivative part
of the current
\begin{eqnarray}
 &  & \mathcal{J}_{s-s_{1}-s_{2}}^{L}=-i\dfrac{\left(s-2\right)!}{\left(2s\right)!}\dfrac{i^{s+s_{1}+s_{2}}\left(1+\left(-1\right)^{s+s_{1}+s_{2}}\right)}{\left(s+s_{1}-s_{2}-1\right)!}\left(s_{1}-1\right)!\left(s_{2}-1\right)!\nonumber \\
 &  & \sum_{d=0}^{s}\left(-1\right)^{d}\left(\begin{array}{c}
s+s_{1}+s_{2}\\
s-d
\end{array}\right)\left(\begin{array}{c}
s-s_{1}-s_{2}\\
d
\end{array}\right)\left(y^{\mu}\right)^{s}\left(\bar{y}^{\dot{\mu}}\right)^{s}\left(D_{\mu\dot{\mu}}\right)^{d}\phi^{\alpha\left(s_{1}\right),\dot{\alpha}\left(s_{1}\right)}\nonumber \\
 &  & \cdot\left\{ \sum_{n=0}^{s_{2}}\left(-1\right)^{n}\left(\begin{array}{c}
s_{2}\\
n
\end{array}\right)\left(D_{\mu\dot{\mu}}\right)^{s-d-n}\left(D_{\alpha\dot{\alpha}}\right)^{s_{1}-s_{2}+n}\phi_{\alpha\left(s_{2}-n\right)\mu(n),\dot{\alpha}\left(s_{2}-n\right)\dot{\mu}(n)}\right\} .
\end{eqnarray}
As in the Section \ref{Max_der}, using Vandermonde's identity (\ref{Vdm})
and integrating by parts one can evaluate the sum over $d$, obtaining
\begin{eqnarray}
 &  & \mathcal{J}_{s-s_{1}-s_{2}}^{L}=-\dfrac{\left(s-2\right)!\left(s_{1}-1\right)!\left(s_{2}-1\right)!}{s!s!\left(s+s_{1}-s_{2}-1\right)!}i^{s+s_{1}+s_{2}+1}\left(1+\left(-1\right)^{s+s_{1}+s_{2}}\right)\left(y^{\mu}\right)^{s}\left(\bar{y}^{\dot{\mu}}\right)^{s}\phi^{\alpha\left(s_{1}\right),\dot{\alpha}\left(s_{1}\right)}\cdot\nonumber \\
 &  & \cdot\left\{ \sum_{n=0}^{s_{2}}\left(-1\right)^{n}\left(\begin{array}{c}
s_{2}\\
n
\end{array}\right)\left(D_{\mu\dot{\mu}}\right)^{s-n}\left(D_{\alpha\dot{\alpha}}\right)^{s_{1}-s_{2}+n}\phi_{\alpha\left(s_{2}-n\right)\mu(n),\dot{\alpha}\left(s_{2}-n\right)\dot{\mu}(n)}\right\} .\label{J_L_fin}
\end{eqnarray}
This completes the analysis of minimal-derivative part of the current.

\subsection{Fronsdal equations with HS current corrections}

Now we are ready to make a final step and write down a current contribution
to quadratic HS equations in Lorentz tensor language. From (\ref{box_eq})
we have

\begin{equation}
\square\phi_{\mu(s),\dot{\mu}(s)}\left(y^{\mu}\right)^{s}\left(\bar{y}^{\dot{\mu}}\right)^{s}+...=-s^{2}\left(s-1\right)\sum_{s_{1}+s_{2}\leq s}\left(\mathcal{J}_{s-s_{1}-s_{2}}^{H}+\mathcal{J}_{s-s_{1}-s_{2}}^{L}\right)+...,\label{box_eq-1}
\end{equation}
where $\mathcal{J}_{s-s_{1}-s_{2}}^{H}$ and $\mathcal{J}_{s-s_{1}-s_{2}}^{L}$
are given in (\ref{J_H_fin}) and (\ref{J_L_fin}) respectively. After
removing twistor variables $y$ and $\bar{y}$, tensor indices are
restored via $\sigma$-matrices, that gives by virtue of
\begin{equation}
\mathrm{Tr}\left\{ \sigma_{a}\bar{\sigma}_{b}\right\} =2\eta_{ab},\quad\left(\sigma^{a}\bar{\sigma}^{b}\sigma^{c}-\sigma^{c}\bar{\sigma}^{b}\sigma^{a}\right)=2i\epsilon^{abcd}\sigma_{d}
\end{equation}
the following result
\begin{eqnarray}
 &  & \square\phi_{a\left(s\right)}+...=\sum_{s_{1}+s_{2}\leq s}\dfrac{\left(s_{1}-1\right)!\left(s_{2}-1\right)!}{\left(s-1\right)!}\left(1+\left(-1\right)^{s+s_{1}+s_{2}}\right)i^{s+s_{1}+s_{2}+1}2^{s_{1}+s_{2}}\cdot\nonumber \\
 &  & \cdot\biggl[\dfrac{\left(-1\right)^{s_{1}}\left(\eta^{2}+\bar{\eta}^{2}\right)}{4\cdot\left(s+s_{1}+s_{2}-1\right)!}\left(D_{a}\right)^{s_{1}-s_{2}}\left(D_{b}\right)^{s_{2}}\phi^{c\left(s_{1}\right)}\cdot\left(D_{a}\right)^{s-s_{1}+s_{2}}\left(D_{c}\right)^{s_{1}}\phi^{b\left(s_{2}\right)}+\nonumber \\
 &  & +\dfrac{\left(-1\right)^{s_{1}}i\left(\eta^{2}-\bar{\eta}^{2}\right)}{4\cdot\left(s+s_{1}+s_{2}-1\right)!}\epsilon_{afcg}D^{f}\left(D_{a}\right)^{s_{1}-s_{2}-1}\left(D_{b}\right)^{s_{2}}\phi^{c\left(s_{1}\right)}\cdot D^{g}\left(D_{a}\right)^{s-s_{1}+s_{2}}\left(D_{c}\right)^{s_{1}-1}\phi^{b\left(s_{2}\right)}+\nonumber \\
 &  & +\dfrac{2^{-s_{2}}}{\left(s+s_{1}-s_{2}-1\right)!}\phi^{b\left(s_{1}\right)}\sum_{n=0}^{s_{2}}\left(-1\right)^{s_{2}+n}\left(\begin{array}{c}
s_{2}\\
n
\end{array}\right)\left(D_{a}\right)^{s-n}\left(D_{b}\right)^{s_{1}-s_{2}+n}\phi_{b\left(s_{2}-n\right)a\left(n\right)}\biggr]+...\label{fin_res_1}
\end{eqnarray}
To simplify the form of this equation one can rescale fields as 
\begin{equation}
\phi_{a\left(n\right)}\longrightarrow\dfrac{2^{-\tfrac{n}{2}}}{\left(n-1\right)!i^{\left(n+1\right)}}\phi_{a\left(n\right)},
\end{equation}
then (\ref{fin_res_1}) turns into 
\begin{eqnarray}
 &  & \square\phi_{a\left(s\right)}+...=\sum_{s_{1}+s_{2}\leq s}\left(1+\left(-1\right)^{s+s_{1}+s_{2}}\right)2^{\tfrac{s+s_{1}+s_{2}}{2}}\nonumber \\
 &  & \biggl\{\cos\left(2\varphi\right)\dfrac{\left(-1\right)^{s_{2}}}{2\cdot\Gamma\left(s+s_{1}+s_{2}\right)}\left(D_{a}\right)^{s}\left(D_{b}\right)^{s_{2}}\phi^{c\left(s_{1}\right)}\cdot\left(D_{c}\right)^{s_{1}}\phi^{b\left(s_{2}\right)}-\nonumber \\
 &  & -\sin\left(2\varphi\right)\dfrac{\left(-1\right)^{s_{2}}}{2\cdot\Gamma\left(s+s_{1}+s_{2}\right)}\epsilon_{afcg}D^{f}\left(D_{a}\right)^{s-1}\left(D_{b}\right)^{s_{2}}\phi^{c\left(s_{1}\right)}\cdot D^{g}\left(D_{c}\right)^{s_{1}-1}\phi^{b\left(s_{2}\right)}+\nonumber \\
 &  & +\dfrac{\left(-1\right)^{s}}{\Gamma\left(s+s_{1}-s_{2}\right)}\sum_{n=0}^{s_{2}}k_{n}\left(D_{a}\right)^{s-n}\phi^{b\left(s_{1}\right)}\cdot\left(D_{b}\right)^{s_{1}-s_{2}+n}\phi_{b\left(s_{2}-n\right)a\left(n\right)}\biggr\}+...\label{fin_res_2}
\end{eqnarray}
where
\begin{equation}
k_{n}:=2^{-s_{2}}\left(\begin{array}{c}
s_{2}\\
n
\end{array}\right),\qquad\sum_{n=0}^{s_{2}}k_{n}=1,
\end{equation}
and we introduced a 'phase angle' $\varphi$
\begin{equation}
\eta=\exp\left(i\varphi\right).
\end{equation}
Let us discuss (\ref{fin_res_2}), which is the main result of the
paper, in some more details. First of all, let us remind that ellipsis
on the l.h.s. denotes the rest of kinetic Fronsdal operator, while
ellipsis on the r.h.s. denotes contributions in $s<s_{1}+s_{2}$ domain
and the contributions of HS currents outside the transverse-traceless
(TT) sector. The non-TT part is completely fixed by the TT one, which
we have found (the procedure of completion of TT part to the full
Lagrangian $AdS$ HS cubic vertex were demonstrated in \cite{SltTar,Francia}).

Next, we see that minimal-derivative part of the current (the last
term in brackets) is $\varphi$-independent, while maximal-derivative
part consists of two different $\varphi$-dependent terms. Term proportional
to $\sin\left(2\varphi\right)$ contains Levi-Civita symbol and thus
is parity-violating, so it expectedly vanishes in parity-invariant
$A$- and $B$-models ($\varphi=0$ and $\varphi=\tfrac{\pi}{2}$).
For parity-invariant models vertices in (\ref{fin_res_2}) coincide
up to a normalisation with the expressions available in the literature
\cite{Metsaev,SltTar}, confirming the correctness of local frame
of Vasiliev equations found in \cite{GelVas_1form}. Another peculiar
situation is $\varphi=\tfrac{\pi}{4}$ model. In this case the first
term with $\cos\left(2\varphi\right)$ is absent, so the maximal-derivative
part of the vertex is in whole proportional to Levi-Civita symbol,
being somewhat of 'maximally parity-breaking'. It would be interesting
to see the implication of that for dual theory, which is conjectured
to be $3d$ Chern-Simons theory coupled to scalar fields \cite{Giombi,Aharony}.

\section{Conclusion}

In the note we obtained quadratic corrections to bosonic Fronsdal
equations generated by gauge-invariant HS currents, starting with
the local second-order Vasiliev equations of \cite{Vas_0form,GelVas_1form}.
The result agrees with previously known expressions \cite{Metsaev,SltTar}
for HS cubic vertices in case of parity-invariant models. This gives
an additional confirmation that the local frame of HS equations, proposed
in \cite{Vas_0form,GelVas_1form}, is the appropriate one. For the
case of $\varphi=\tfrac{\pi}{4}$ model we found that maximal-derivative
part of the vertex is proportional to Levi-Civita symbol, being maximally
parity-breaking, that may have interesting consequences for dual boundary
Chern-Simons theory. It would be interesting also to study the theories
with fermions as well as to find the contribution of gauge-dependent
sector, that would allow one to write down the full quadratic HS equations.

\section*{Acknowledgements}

The author is grateful\textbf{ }to O.A. Gelfond for important technical
remark, to V.E. Didenko for useful comments, to D. Francia and M.
Taronna for the correspondence, and especially to M.A. Vasiliev for
numerous valuable discussions and critical remarks. The work was partially
supported by the Russian Basic Research Foundation Grant No. 17-02-00546
and by the Foundation for Theoretical Physics Development ``Basis''.

\end{document}